\documentclass[twocolumn]{svjour3}          % twocolumn
\smartqed  % flush right qed marks, e.g. at end of proof
\usepackage{graphicx}
\begin{document}
\title{Disparities of Larmor's/Li\'{e}nard's formulations with special relativity and energy-momentum conservation}
\titlerunning{Disparity of Larmor's formula with energy-momentum conservation}  
\author{Ashok K. Singal}
\authorrunning{A. K. Singal} 
\institute{A. K. Singal \at
Astronomy and Astrophysics Division, Physical Research Laboratory,
Navrangpura, Ahmedabad - 380 009, India\\
\email{ashokkumar.singal@gmail.com}         }
\date{Received: date / Accepted: date}
\maketitle
\begin{abstract}
It is shown that the familiar Larmor's formula or its relativistic generalization, 
Li\'{e}nard's formula, widely believed to represent the instantaneous radiative losses from an accelerated charge, are not compatible with the special theory of relativity (STR). It is also shown that energy-momentum  conservation is violated in all inertial frames when we compute the instantaneous rate of loss of kinetic energy and momentum of the radiating charge in accordance with Larmor's/Li\'{e}nard's formulation. It is emphasized that one should clearly distinguish between the electromagnetic power going as radiation into space 
and the instantaneous loss of mechanical power by the charge. In literature both powers are treated as not only equal but almost synonymous; the two need not be the same however. It is pointed out that 
a mathematical subtlety in the applicability of Poynting's theorem seems to have been overlooked in the text-book derivation of Larmor's formula, where a proper distinction between ``real'' and ``retarded'' times was not maintained. This has led to some century-old apparent anomalies, including the mysterious Schott energy, which happens to be nothing but merely a change in self-field energy of the charge between retarded and real times.
\end{abstract}
%--------------------------
\section{Introduction}
Larmor's formulation (or its relativistic generalization, Li\'{e}nard's result) is generally believed to yield instantaneous radiative power losses by an accelerated charge where the radiating charge apparently loses 
its kinetic energy into electromagnetic radiation at a rate proportional to the square of its acceleration \cite{la97,1,2,25}. One can also compute the rate of momentum carried by the electromagnetic radiation and exploiting the momentum conservation one can then calculate the resulting radiation damping force on the charge. This damping force turns out to be directly proportional to the velocity vector of the charge multiplied by the power being radiated \cite{ha95,51}.

In an alternate approach, assuming the charge to be a small sphere, one can obtain the net self-force on the accelerated charge, calculated in a rather cumbersome way from the detailed mutual interaction between its constituents, evaluated usually in its instantaneous rest frame. This force, widely known in literature as radiation reaction, turns out to be  proportional to the rate of change of acceleration of the charge, independent of the assumed radius of the charged sphere \cite{lor04,abr04,abr05,16}. 
This radiation reaction formula can also be derived directly from the rate of electromagnetic momentum flow, calculated using the Maxwell stress tensor, across a surface surrounding the  neighbourhood of a point charge \cite {68b}.
One then gets an instantaneous radiative power loss formula from the rate of work done against the self-force of the accelerated charge, by a scalar product of the self-force with the velocity of the charge. The same formula for the radiative power loss is also obtained independently from the Poynting flux in the neighbourhood of a point charge in arbitrary motion \cite {68a}.

The dissimilarity between power loss given by Larmor's formula (proportional to the square of acceleration) and that given by the radiation reaction (proportional to the scalar product of the rate of change of acceleration 
with the velocity of the charge) has remained an enigma for more than a century. Difference in the two power loss formulas appears as a ``Schott'' term (because it was first pointed out by Schott \cite{7}), which is thought to represent an extra energy term in fields of an accelerated charge. 
However, the genesis of this acceleration-dependent energy term is still not clear \cite{51,6,44,eg06,41,56,row10} and it does not seem to make an appearance elsewhere in physics. 

The general consensus is that it is Larmor's  formula (or its relativistic generalization, Li\'{e}nard's formula), that begets correct results \cite{51,7,23,26,27}. 
Here we shall demonstrate the incompatibility of Larmor's and of Li\'{e}nard's formulation with the STR by analysing a charge undergoing synchrotron losses. An advantage in this particular approach is that the agency responsible for the acceleration of the charge, viz. the magnetic field, does no work as the force caused by it is always perpendicular to the instantaneous velocity of the charge. Thus there is no ambiguity here in that the power losses against the electromagnetic radiation have to be undergone by the charge itself and not by the agency responsible for the 
acceleration. We shall show that the energy-momentum losses due to radiating damping, as 
calculated in two different inertial frames, are not compatible with the relativistic 
transformation laws. Further, the energy-momentum conservation is violated in all 
inertial frames when we employ Larmor's/Li\'{e}nard's formulation to calculate the loss of kinetic energy and momentum of the accelerated charge, or where appropriate, that of the agency responsible for the acceleration. 
Here we pin down the root cause of these anomalies to a mathematical subtlety in the applicability of  Poynting's theorem that has been overlooked in the text-book derivation of Larmor's formula. 
Further, we point out that the mysterious Schott term is nothing but the difference in the rate of change of energy in self-fields of the charge between retarded and real times.

%-----------------------------------------------------------
\section{The standard picture}
Larmor's formula for electromagnetic power radiated from an accelerated charge in its instantaneous rest frame ${\cal K}'$ is \cite{la97,1,2,25}
\begin{equation}
\label{eq:21a}
{{\cal P}'_{\rm rad}}=\frac{2e^2}{3c^3}\:[\dot{\bf v}'\cdot\dot{\bf v}'] \:.
\end{equation}
In our formulations, we throughout use Gaussian (cgs) system of units. 

Since the radiated power of the accelerated charge in its instantaneous rest frame ${\cal K}'$ has an azimuth symmetry ($\propto \sin^2\phi'$) \cite{1,2,25}, the momentum carried away by the radiation in one hemisphere, say $\phi'=0$ to $\pi$, cancels that in the opposite hemisphere ($\phi'=\pi$ to $2\pi$), implying that an integration over the solid angle yields a zero value for the net rate of momentum loss. 
\begin{equation}
\label{eq:21a1}
\dot{\bf p}'_{\rm rad}=0\:.
\end{equation}
Consequently, the force on the charge due to radiation damping is zero in its instantaneous rest frame.

Now $\gamma {\cal P}_{\rm rad}/c, \gamma\dot{\bf p}_{\rm rad}$ form a 4-vector under relativistic transformation \cite{1,71,mo94}. Then we can write
\begin{equation}
\label{eq:21a2}
\gamma {\cal P}_{\rm rad}=\gamma ({\cal P}'_{\rm rad}+\dot{\bf p}'_{\rm rad}\cdot{\bf v}/c) \:,
\end{equation}
With the force due to radiation damping being nil in the rest frame $(\dot{\bf p}'_{\rm rad}=0) $, we have 
\begin{equation}
\label{eq:21a3}
{\cal P}_{\rm rad}={\cal P}'_{\rm rad} \:,
\end{equation}
implying that the radiated power is an invariant.

The parallel and perpendicular components of acceleration respectively transform from the rest frame ${\cal K}'$ to the lab-frame ${\cal K}$ as \cite{71}
\begin{equation}
\label{eq:3f2.1a}
\dot{ v}_\parallel = \frac{\dot{ v}'_\parallel}{\gamma^3} 
\end{equation}
\begin{equation}
\label{eq:3f2.1b}
\dot{ v}_\perp = \frac{\dot{ v}'_\perp}{\gamma^2} \:.
\end{equation}
Then we straightway get Li\'{e}nard's formula for power radiated from a charge moving with an arbitrary velocity.
\begin{eqnarray}
\nonumber
{\cal P}_{\rm rad}&=&\frac{2e^{2}\gamma ^{4}}{3c^{3}}\left[\gamma ^{2}\dot{ v}_\parallel^{2}+\dot{ v}_\perp^{2}\right]\\
\label{eq:11}
&=&\frac{2e^{2}\gamma ^{6}}{3c^{3}}\left[\dot{\bf v}^{2}-
\frac{({\bf v}\times\dot{\bf v})^{2}}{c^{2}}\right].
\end{eqnarray}

A more formal expression for the 4-vector representing the rate of energy-momentum being carried away by radiation from the charge is
\begin{equation}
\label{eq:21b1}
{\cal F}^{\mu}=\frac{2e^{2}}{3c^{5}} \:\dot{v}^{\alpha}\dot{v}_{\alpha}{v}^{\mu} \:,
\end{equation}
where  in this and other covariant equations, dot denotes differentiation with respect to proper time of the charge \cite{51}. ${\cal F}^{0}$ is nothing but $\gamma {\cal P}_{\rm rad}/c$, where
\begin{equation}
\label{eq:11b}
{\cal P}_{\rm rad}=\frac{2e^{2}}{3c^{3}} \:\dot{v}^{\alpha}\dot{v}_{\alpha}
%=\frac{2e^{2}\gamma ^{4}}{3c^{3}}\left[\dot{\bf v}\cdot\dot{\bf v} +
%\left(\frac{\gamma\dot{\bf v} \cdot{\bf v}}{c}\right)^2\right]
\end{equation} 
is Li\'{e}nard's formula (Eq.~(\ref{eq:11})) for power going into radiation from  a charge moving relativistically, 
and ${\cal F}^{i}$ ($i=1,2,3$) is $\gamma$ times the $i$th component of the rate of momentum being carried away by the radiation
\begin{equation}
\label{eq:21b}
\dot{\bf p}_{\rm rad}=\frac{{\cal P}_{\rm rad}}{c^2}{\bf v}\:.
\end{equation}

According to the conventional wisdom \cite{ha95,51}, from energy-momentum conservation, kinetic energy of the charge should change due to radiation losses at a rate 
\begin{equation}
\label{eq:11a}
\dot{\cal E}=-{\cal P}_{\rm rad},
\end{equation}  
while the charge would experience a force due to radiation damping  
\begin{equation}
\label{eq:21b2}
{\bf F}=-\dot{\bf p}_{\rm rad}=-\frac{{\cal P}_{\rm rad}}{c^2}{\bf v}\:.
\end{equation}
From Eq.~(\ref{eq:21b2}), the rate of change of momentum of the charge due to radiation damping, ought to be in a direction opposite to its instantaneous velocity vector \cite{51}. This immediately implies that while the kinetic energy and hence the speed of the charge will reduce due to radiation losses, its direction of motion would not be affected \cite{23,26,27}. 
We shall demonstrate that this inference is not commensurate with the STR and that depending upon individual cases, the direction of motion of the charge may also undergo a change because of radiation damping. 

%--------------------------------------
\section{Incongruity of synchrotron losses in the standard picture with special relativity}
Let us consider the Lab-frame ${\cal K}$ in which there is a uniform magnetic field $B$ along the z-axis 
(Fig.(1)). 
A relativistic charged particle, say, an electron, 
having a velocity ${\bf v}$ and Lorentz factor $\gamma=1/\sqrt{1-v ^{2}/c^2}$ 
moves in a helical path with $\theta$ as the pitch angle, defined  as the angle that the velocity 
vector makes with the magnetic field direction. 
From the standard electromagnetic radiation theory \cite{1,2,25,26,27}, the radiation from a relativistic charge lies predominantly within a narrow cone of angle $\sim 1/\gamma$ around the instantaneous forward direction of the charge motion and the momentum carried by the radiation (Eq.~(\ref{eq:21b})) would be along the instantaneous direction of motion of the charge. From conservation of momentum it can then be construed that the radiation damping would 
cause only a decrease in the magnitude of the velocity vector of the charge without affecting its direction. 
Thus, according to the conventional wisdom, the ratio $v_\perp/v_\parallel= \tan\theta$, should not change. 
%-------------------------------------------
\begin{figure}
\includegraphics[width=\columnwidth]{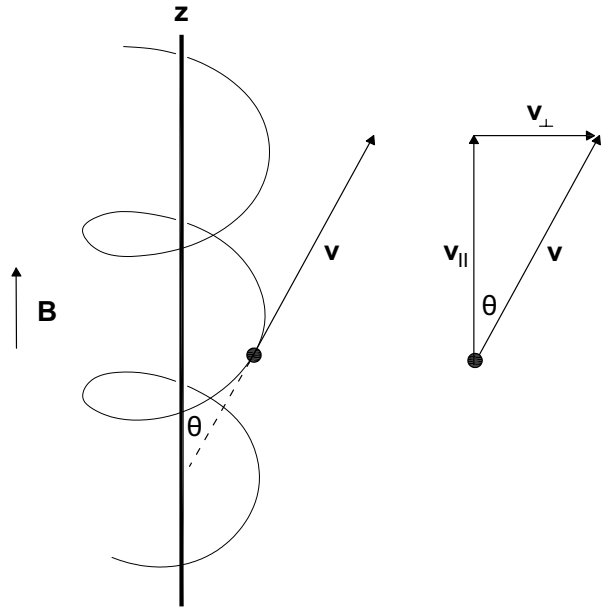}
\caption{As seen in the Lab-frame ${\cal K}$, the electron moves in a helical orbit with a velocity $\bf v$ along a pitch angle 
$\theta$ in a uniform magnetic field $\bf B$. $\bf v_{\parallel}$ is the velocity component parallel to 
the magnetic field along the $z$-axis while $\bf v_{\perp}$, the perpendicular velocity component, 
lies in the $x$-$y$ plane.} 
\end{figure}
%-------------------------------------------

As a result, pitch angle $\theta$ of the motion should not change due to radiative losses in a synchrotron process. The accordingly derived formulas for energy losses with constant pitch angles of synchrotron electrons in radio galaxies are the standard text-book 
material for the last 50 years \cite{23,26,27}. However, using the velocity transformation from special relativity, we show below that the pitch angle of a radiating charge varies. While the velocity 
component $v_\parallel$, parallel to the magnetic field, remains unaffected, the perpendicular component $v_\perp$ does 
reduce in magnitude due to radiative losses, implying a change in the pitch angle.

As there is no force component due to the magnetic field parallel to itself, a charge with a velocity component 
$v_{\parallel}$ keeps on moving along the $z$-axis with a constant velocity, unaffected by the field. However, we want to examine the possibility whether the force due to radiation damping might affect this velocity component. From the standard  arguments in the literature \cite{23,26,27}, due to radiation damping the speed of the charge will reduce without a corresponding change in its direction of motion, 
therefore a charge moving with a finite pitch angle should suffer proportional losses in $v_{\parallel}$ and $v_\perp$, so as to keep the pitch angle constant.  
It means that due to radiative losses the charge would attain an acceleration which has a component, $\dot{v}_{\parallel}$  as well, along the $-z$ direction.
However, as we will show from the theory of relativity, such a charge cannot attain an acceleration parallel to the magnetic field and 
that there is something amiss in these above arguments that needs a careful scrutiny. 

%----------------------------------------------------------------
\subsection{A PARADOX!}
Let us consider an inertial frame ${\cal K}''$ which moves with a velocity $\bf v_{\parallel}$ with respect to the Lab-frame ${\cal K}$. Since the magnetic field in frame ${\cal K}$ is along the $z$-axis and the relative velocity of ${\cal K}''$ with respect to ${\cal K}$ is also along the $z$-axis,  the electromagnetic field transformation between ${\cal K}$ and ${\cal K}''$ will leave the magnetic field 
unchanged. Moreover, $\bf v_{\parallel}\times \bf B=0$, implies no accompanying electric field in frame ${\cal K}''$. The charged particle will have no velocity component parallel to the magnetic field in ${\cal K}''$ (${ v}''_{\parallel}=0$) and will have only a circular motion in a plane perpendicular to the magnetic field (with a pitch angle $\theta''=\pi/2$). The gyro center of the charge will be stationary in  frame ${\cal K}''$, which is therefore often referred to as the gyro-center (GC) frame. Due to radiative losses by the charge, there will be a decrease in speed along the velocity vector which, as seen in the frame ${\cal K}''$, is solely  
in a plane perpendicular to the magnetic field. The charge will thus continue to gyrate in a circle of a gradually decreasing radius in a plane perpendicular to the magnetic field, without ever getting any  velocity component along the  direction of the magnetic field. 
%----------------------------------------
Thus in the GC frame ${\cal K}''$, there is no component of radiation damping force or acceleration parallel to the magnetic field, i.e., ${ f}''_\parallel = 0$, $\dot{ v}''_\parallel = 0$. 

A relativistic transformation of velocity between ${\cal K}$ and ${\cal K}''$ gives \cite{71} 
\begin{equation}
\label{eq:3f2a}
{ v}_\perp =  \frac{{ v}''_\perp}{\gamma_\parallel} \:,
\end{equation} 
where $\gamma_\parallel=1/\sqrt{1- v_\parallel^2/c^2}$ is the Lorentz factor corresponding to the velocity $\bf v_{\parallel}$ between ${\cal K}$ and ${\cal K}''$.

Now, in the GC frame ${\cal K}''$, the force $\bf F''$ due to radiation damping ($\propto \bf v''$; Eq.~(\ref{eq:21b2})) 
is always in a plane perpendicular to the magnetic field.
From the 4-force transformation \cite{71,mo94} to the lab frame ${\cal K}$, we have
\begin{equation}
\label{eq:10.1.6}
{F}_\parallel=\frac{{\cal P}{ v}_\parallel}{c^2}=\frac{({\bf F''\cdot { v}''}){ v}_\parallel}{c^2}=\frac{F''_\perp { v}''_\perp{ v}_\parallel}{c^2}=\frac{F''_\perp { v}_\perp{ v}_\parallel\gamma_\parallel}{c^2},
\end{equation}
\begin{equation}
\label{eq:10.1.7}
{F}_\perp=\frac{F''_\perp}{\gamma_\parallel}\;.
\end{equation}
We note that the ratio of the force components is not equal to that of the velocity components 
\begin{equation}
\label{eq:10.1.3a}
\frac{{F}_\perp}{{F}_\parallel}=\frac{1}{(\gamma_\parallel^2 v_\parallel v_\perp)}\ne  \frac{v_\perp}{ v_\parallel}\;.
\end{equation}
implying that the force vector due to radiation damping is not parallel to the velocity vector, which thus is inconsistent with Eq.~(\ref{eq:21b2})

A curious thing to note here is that even though  in frame ${\cal K}$ a finite parallel force component exists (${F}_\parallel \ne 0$; Eq.~(\ref{eq:10.1.6})), there is nil acceleration component along the $z$-axis, i,e,  $\dot{v}_\parallel=0$. 
Actually in frame ${\cal K}$, a force component along $z$ direction shows up solely because of 
a finite rate of change of $\gamma$ due to $\dot{v}_{\perp}$, even though $\dot{v}_{\parallel}=0$.
It may be recalled that in special relativity, force and acceleration vectors are not
always parallel, e.g., in a case where force is not parallel to the velocity vector, the acceleration need not be 
along the direction of the force. When the applied force is either parallel or perpendicular to the velocity vector, 
only then is the acceleration along the direction of the force \cite{71}. 

In the present case this assertion can be verified from a relativistic transformation of acceleration components
between ${\cal K}''$ and ${\cal K}$ \cite{71}.
\begin{equation}
\label{eq:3f2.2a}
\dot{ v}_\parallel =  \frac{\dot{ v}''_\parallel}{\gamma_\parallel^3} \:,
\end{equation}
\begin{equation}
\label{eq:3f2.2b}
\dot{ v}_\perp =  \frac{\dot{ v}''_\perp - \dot{ v}''_\parallel { v}''_\perp{ v}_\parallel/c^2}{\gamma_\parallel^2}\:,
\end{equation}
It is imperative to note that ${\cal K}''$ is not a rest frame of the charge and the expression for transformation of $\dot{ v}_\perp$ in  Eq.~(\ref{eq:3f2.2b}) differs from that in Eq.~(\ref{eq:3f2.1b}). 
Further, it should be kept in mind that the acceleration in question  
here is not that due to the force by the magnetic field on the moving charge (which is perpendicular to the 
instantaneously velocity of the charge), but the acceleration (or rather a deceleration) caused on the charge due to  
radiation damping.

Since $\dot{v}''_{\parallel}=0$,  
we have 
\begin{equation}
\label{eq:3f2.3a}
\dot{ v}_\parallel =  0
\end{equation}
\begin{equation}
\label{eq:3f2.3b}
\dot{ v}_\perp =  \frac{\dot{ v}''_\perp}{\gamma_\parallel^2}\:,
\end{equation}
%-------------------------------------------------

In physical terms, since in the GC frame ${\cal K}''$, the charge does not attain any motion along  
the magnetic field ($\dot{\bf v}''_{\parallel}=0$ always), and the two inertial frames 
(${\cal K}$ and ${\cal K}''$) have a constant relative motion, the parallel component of velocity, $v_\parallel$, of the charge should remain constant in ${\cal K}$ too. As the magnitude 
of the perpendicular component $v''_\perp$ constantly diminishes in ${\cal K}''$ because of radiative damping,
$v_\perp$ in ${\cal K}$ as well would be decreasing. Therefore the pitch angle $\tan\theta = v_\perp/v_\parallel$ of the charge, as seen in ${\cal K}$, should also decrease with time. Consequently, velocity of the charge increasingly aligns with the magnetic field direction.
Accordingly we arrive at conflicting results. While an application of radiation damping (Eq.~(\ref{eq:21b2})), calculated from Larmor's formula in inertial frame ${\cal K}$, led us to the result that 
the pitch angle of the charge is a constant, a similar application of the same equation in the GC frame ${\cal K}''$, followed by a relativistic transformation to frame ${\cal K}$, begets  
results that violate Eq.~(\ref{eq:21b2}) leading to a conclusion that the pitch angle will be progressively reducing in ${\cal K}$ as the charge radiates with time. 

Thus we encounter a {\bf paradox}. 
The two approaches, one using Larmor's formula directly and the other using Larmor's formula in conjunction with the STR, lead to two different results for the same physical quantity, the pitch angle. Thus we find that the predictions of the radiation damping according to the Larmor's formula \cite{ha95,51}, is not in accordance with the STR.

%----------------------------------------------------
\section{Inconsistency with energy-momentum conservation}
Larmor's formulation is inherently inconsistent when applied to compute loss of energy-momentum of a charge through radiation. Let us consider the simple case of a one-dimensional motion with $\dot{\bf v} \parallel {\bf v}$. Then from Eqs.~(\ref{eq:11a}) and (\ref{eq:21b2}) we can write
\begin{equation}
\label{eq:21c}
\dot{\cal E}= m \gamma^3\dot{v}{v}= -{\cal P}_{\rm rad}\:.
\end{equation}
\begin{equation}
\label{eq:21d}
{F}= m \gamma^3 \dot{v}=-\frac{{\cal P}_{\rm rad}}{c^2}{v}\:.
\end{equation}
Equations (\ref{eq:21c}) and (\ref{eq:21d})) are mutually inconsistent for any value of ${v}$. Actually this inconsistency is apparent from Eq.~(\ref{eq:21a}) itself as 
in the rest frame of the charge, the energy loss rate is finite even when the charge 
has no kinetic energy to lose. Also from Eq.~(\ref{eq:21a1}), the net momentum carried away by the radiation is zero, 
thus in any case the charge would have to lose kinetic energy without losing any momentum.  Therefore energy-momentum conservation is clearly violated and one cannot get consistent results, using Larmor's 
or its relativistic generalization, Li\'{e}nard's formula, to calculate the radiative power losses and the radiation damping force on the accelerated charge. 

One could argue that it may be the external agency responsible for accelerating the charge, for instance, an electric field due to some external source, that might actually provide the force and power to compensate for the radiation losses as well.
\begin{equation}
\label{eq:21d1}
f_{\rm ext}= m \gamma^3 \dot{v} + \frac{{\cal P}_{\rm rad}}{c^2}{v}\:.
\end{equation}\begin{equation}
\label{eq:21c1}
f_{\rm ext}\:{v}= m \gamma^3\dot{v}{v} + {\cal P}_{\rm rad}\:.
\end{equation}
But Eqs. (\ref{eq:21d1}) and (\ref{eq:21c1}) are still mutually inconsistent for any value of ${v}$.

Actually, the only way to make Eqs.~(\ref{eq:21a}) and (\ref{eq:21a1}) to conform to the  energy-momentum conservation is if the charge were converting its rest-mass energy into radiation, because only in that case could one have 
energy loss in the rest frame without any accompanying loss in momentum. Even in the Lab-frame ${\cal K}$,  Eqs.~(\ref{eq:11a}) and (\ref{eq:21b2}) can hold true simultaneously only if the charge were to lose its rest mass, without affecting its speed or relativistic Lorentz factor $\gamma$, giving us 
\begin{equation}
\label{eq:21e}
\dot{\cal E}= \dot{m}{c^2} \gamma = -{\cal P}_{\rm rad}\:,
\end{equation}
\begin{equation}
\label{eq:21f}
{F}=\dot{m}\gamma {\bf v}=-\frac{{\cal P}_{\rm rad}}{c^2}{\bf v}\:,
\end{equation}
which seem consistent.

Accordingly, in Larmor's formula (or its relativistic generalization, Li\'{e}nard's radiation formula), at its face value, the total energy of the radiating 
charge would certainly decrease, but not its speed or the Lorentz factor $\gamma$. It is the rest mass of the charge that would get converted into radiation energy. Such an idea has actually been  proposed in past that the {\em proper mass} of an accelerated charged particle varies \cite{8}. In such a scenario, the rate of change of momentum would be due to the decrease in the rest mass, without implying any change in the velocity vector, and the power loss will be equal in all frames.
Then only the formula 
for radiation damping derived from Larmor's formula for radiation and widely used in literature, will yield consistent results in all inertial frames, 
including the Lab-frame  ${\cal K}$ as well as the GC frame  ${\cal K}''$, discussed earlier in the synchrotron radiation case.

However, if we want to discard this seemingly preposterous idea of the rest mass of the accelerated charge getting converted into electromagnetic radiation and instead take the much more preferable  view  that an electron (or for that matter any radiating charge) does not lose its rest mass during the radiation process and internally remains still the same and that the radiative losses are from the kinetic energy of the radiating charge, then there is no way that Larmor's formula (Eqs.~(\ref{eq:11a}) and (\ref{eq:21b2})) could be applied to calculate {\em instantaneous energy-momentum losses}  of a radiating charge.

%-----------------------------------------------------
\section{Wherein lies the fallacy? -- Real time versus retarded time}
It is rather consternating that a century-old, widely used formula should harbour such anomalies. 
The electromagnetic power going into radiation from an accelerated charge is derived after all by a legitimate text-book calculation of Poynting flux through a spherical surface of large enough radius, $R\rightarrow \infty$. The usual assumption is that for the radiated power computation, it is only the acceleration fields ($\propto \dot{v}/R c^2$) that matter \cite{1,2,25} as the contribution of velocity fields falls rapidly with distance ($\propto v/R^2c$). Only exception perhaps is in the case of a uniformly accelerated charge, where the velocity depends on the retarded time as $v \propto \dot{v} (t'-t) = -\dot{v} R/c$ and consequently the expression for velocity fields also comprises a term $\propto -\dot{v}/Rc^2$, which could then match and thereby even cancel the contribution of acceleration fields \cite{17,18,68c,68e}.
Barring such exceptions, Larmor's formula does yield the electromagnetic power being radiated to infinity. 

However, one cannot legitimately use Poynting's theorem to equate this radiation power at time $t$ to the rate of loss of the mechanical energy $({\cal E_{\rm me}})$ of the charge at a retarded time $t'=t-R/c$, as usually done in the textbook derivations \cite{1,2,25}, but which, strictly speaking, is not valid.
\begin{equation}
\label{eq:23b}
\left[\frac{{\rm d}{\cal E_{\rm me}}}{{\rm d}t}\right]_{t'}
\ne
- \left(\int_{\Sigma}{{\rm d}\Sigma}\:({\bf n} \cdot {\cal S})\right)_{t} \;.
\end{equation}

Even though the electromagnetic fields at $R$ at time $t$ do 
get determined by the charge dynamics at the retarded time $t'=t-R/c$, Poynting's theorem, however, does not provide a direct relation between the Poynting flux at $t$ and the mechanical power loss of the charge at $t'$ (Eq.~(\ref{eq:23b})). 
It is this oversight which is mostly responsible for the confusion in this century old controversy. Poynting theorem strictly relates the Poynting flux through  
a surface enclosing a volume, to the rate of loss of the mechanical energy $({\cal E_{\rm me}})$ of the charge and that 
of the energy of the electromagnetic fields $({\cal E_{\rm em}})$ within the volume, all evaluated 
{\em for the same instant of time} $t$. 
\begin{equation}
\label{eq:23a}
\left(\frac{{\rm d}{\cal E_{\rm me}}}{{\rm d}t}+\frac{{\rm d}{\cal E_{\rm em}}}{{\rm d}t}\right)_{t}
=-\left(\int_{\Sigma}{{\rm d}\Sigma}\:({\bf n} \cdot {\cal S})\right)_{t} \;.
\end{equation}
A correct application of the Poynting's theorem, in terms of the charge motion at real time $t$, leads to a formula for  instantaneous {\em power loss} by the charge (in a non-relativistic motion) as \cite{68a}
\begin{equation}
\label{eq:6.1}
{\cal P} =-\frac{2e^{2}}{3c^{3}}\ddot{\bf v}\cdot{\bf v}\:,
\end{equation}
while a similar real-time application of the momentum conservation theorem, employing the Maxwell's stress tensor, yields a radiation drag force on the charge \cite{68b} 
\begin{equation}
\label{eq:3a}
{\bf f}= \frac{2e^{2}}{3c^{3}}\ddot{\bf v}\:.
\end{equation}
Equation~(\ref{eq:3a}) is the well-known Abraham-Lorentz radiation reaction formula \cite{lor04,abr04,abr05,16,7,20},  derived usually in a quite cumbersome way from the net force on the charge from its self-fields at the retarded time, but in  \cite{68b} it was obtained first time from the  momentum conservation theorem. It has been explicitly shown \cite{31,48} that the radiation reaction given by Eq.~(\ref{eq:3a}) (or rather its relativistic generalization) yields results consistent with STR for the synchrotron radiation losses.

As we mentioned earlier, it is a general belief that the instantaneous radiative losses from a charge are given by Larmor's formula (Eq.~(\ref{eq:21a})). The difference between power loss formulas given by Eqs.~(\ref{eq:6.1}) and (\ref{eq:21a}) is 
\begin{equation}
\label{eq:9.1}
-\frac{2e^{2}}{3c^{3}}\ddot{\bf v}\cdot{\bf v}-\frac{2e^{2}}{3c^{3}}\dot{\bf v}\cdot\dot{\bf v} 
=-\frac{2e^{2}}{3c^{3}}\frac{{\rm d}(\dot{\bf v}\cdot{\bf v})}{{\rm d}t}\:.
\end{equation}
The term on the right hand side in Eq.~(\ref{eq:9.1}) is known as the Schott term and 
it has been recently demonstrated that the Schott term represents the difference in rate of change of energy in self-fields of the charge between the retarded and the real times \cite{68c,68} and  contrary to the ideas prevalent in the literature  \cite{51,6,44,eg06,41,56,row10} it is not an acceleration-dependent extra energy present in fields, as has been explicitly demonstrated by a detailed scrutiny of the electromagnetic fields, in the case of a uniformly accelerated charge \cite{68e}.

It is important to note that though the formulation for the electromagnetic power radiated into far-off space remains unaltered from 
that given in the text-books (except perhaps in the cases where contribution of velocity fields could not be neglected, for example, for a uniformly accelerated charge \cite{18,68c,68e}) Larmor's formula or its relativistic generalization Li\'{e}nard's result, however, cannot be used to 
infer {\em instantaneous radiative losses} by the charge. The latter have to be instead inferred from Eq.~(\ref{eq:6.1}) or its relativistic generalization \cite{68b}
\begin{equation}
\label{eq:10}
{\cal P}= -\frac{2e^{2}\gamma ^{4}}{3c^{3}}\left[\ddot{\bf v}\cdot{\bf v}+
3\gamma ^{2}\frac{(\dot{\bf v}\cdot{\bf v})^{2}}{c^{2}}\right]\:,
\end{equation}
which alone yields a correct value for the instantaneous radiative losses.  

It is therefore, imperative that one treats the electromagnetic power going into far-off space and the instantaneous loss of mechanical power by the charge as two distinct quantities. In literature both powers are considered not only equal but almost identical. However, the two could be quite different.
The disparity with the special theory of relativity or inconsistency with energy-momentum conservation laws manifests itself only when we try to use Larmor's (or Li\'{e}nard's) formula for calculating instantaneous energy-momentum loss of the radiating charge. It ought to be now clear why it happens so. Larmor's formula gives radiated power in terms of charge motion at the retarded time, and as a result does not take into account the variation in the self-field energy as the charge velocity changes due to acceleration between the retarded and real times. On the other hand, the power loss calculated from the Abraham-Lorentz radiation reaction formula yields only the excess power going into fields over and above the real-time change taking place in the self-field energy, thereby correctly representing the radiative {\em losses}. 

We may add here, 
however, that everything is not lost as Larmor's formula may yet give correct result in most cases
for the {\em average} power loss \cite{68a}, even though the strict instantaneous rate could be very different. This is so because a time average of the Schott term over a complete period would yield a nil value in a periodic motion, 
and almost any actual motion of the charge could be Fourier analysed as a sum of periodic components, one conspicuous exception being the motion of a uniformly accelerated charge.

\section{Conclusions}
It was demonstrated that the radiation damping, as given by Larmor's formula or its relativistic generalization, 
Li\'{e}nard's formula, is not commensurate with the special theory of relativity. It was also shown that energy-momentum  conservation is violated if we compute the instantaneous rate of loss of kinetic energy and momentum of the radiating charge in accordance with Larmor's/Li\'{e}nard's formulation. Though Larmor's formula gives in almost all cases the electromagnetic power going as radiation into space, but it does not unequivocally represent an instantaneous loss of mechanical power by the charge, 
yet it may give the power loss in a time-averaged sense. We brought out the 
underlying cause in the genesis of the problem which is  
a mathematical subtlety in the applicability of Poynting's theorem that seems to have been overlooked in the text-book derivation of Larmor's formula, where a proper distinction between ``real'' and ``retarded'' times was not maintained. We argued that one should instead use the Abraham-Lorentz radiation reaction formula (or 
its relativistic generalization when appropriate) to calculate the instantaneous rate of energy loss by the radiating charge. Further, it was pointed out 
that the Schott energy term does not denote any acceleration-dependent extra energy but merely represents a  change in self-field energy of the charge between retarded and real times.
%-------------------------------------------
\section*{Acknowledgements} 
I acknowledge Apoorva Singal for her help in the preparation of diagram. 
%--------------------
%--------------------
{}
\end{document}